\documentclass[pdflatex,sn-mathphys-num]{sn-jnl}

\usepackage{graphicx}%
\usepackage{multirow}%
\usepackage{amsmath,amssymb,amsfonts}%
\usepackage{amsthm}%
\usepackage{mathrsfs}%
\usepackage[title]{appendix}%
\usepackage{xcolor}%
\usepackage{textcomp}%
\usepackage{manyfoot}%
\usepackage{booktabs}%
\usepackage{algorithm}%
\usepackage{algorithmicx}%
\usepackage{algpseudocode}%
\usepackage{listings}%

\begin{document}

\title[GINsim 3.1]{GINsim 3.1: Qualitative modelling and dynamical analysis of logical regulatory networks}



\author[1]{\fnm{Laurent} \sur{Tichit}}\email{laurent.tichit@univ-amu.fr}

\author[2]{\fnm{Dominique} \sur{Benielli}}\email{dominique.benielli@univ-amu.fr}

\author[1]{\fnm{Olivier} \sur{Chabrol}}\email{olivier.chabrol@univ-amu.fr}

\author*[3,4]{\fnm{Denis} \sur{Thieffry}}\email{denis.thieffry@bio.ens.psl.eu}

\author[1]{\fnm{Claudine} \sur{Chaouiya}$^{\dagger}$}

\author*[5]{\fnm{Pedro T.} \sur{Monteiro}}\email{pedro.tiago.monteiro@tecnico.ulisboa.pt}

\affil[1]{\orgname{Aix Marseille University, CNRS, I2M}, \orgaddress{\city{Marseille}, \country{France}}}

\affil[2]{\orgname{Aix Marseille University}, \orgaddress{\city{Marseille}, \country{France}}}

\affil[3]{\orgname{Institute of Biology of the École Normale Supérieure, CNRS, INSERM}, \orgaddress{\city{Paris}, \country{France}}}

\affil[4]{\orgname{Institut Curie, MinesParis, INSERM}, \orgaddress{\city{Paris}, \country{France}}}

\affil[5]{\orgname{INESC-ID / Instituto Superior Técnico, University of Lisbon}, \orgaddress{\city{Lisbon}, \country{Portugal}}}

{\def\thefootnote{\relax}\footnotetext{$^{\dagger}$Claudine Chaouiya (1961-2025) co-directed the GINsim project since its initiation in 2002 and strongly pushed the development of this new release, before she passed away in September 2025. Therefore, we decided to keep Claudine as co-last author and to dedicate this manuscript to her memory.}}


\abstract{
GINsim is a software for the definition and dynamical analysis of qualitative (Boolean and multi-valued) models of biological regulatory networks.
Building on its modular architecture, based on the MDDlib decision-diagram and bioLQM model-manipulation libraries, GINsim version 3.1 introduces new functionalities aimed at the analysis of large logical models: the symbolic computation of trap-spaces, the Booleanisation of multilevel models and the reversal of model dynamics.
GINsim 3.1 also extends the estimation of attractor reachability (Monte Carlo and Avatar algorithms), the computation of stable states, as well as model reduction functionalities.
Analyses can be applied to perturbed models, and scripted, in particular from the CoLoMoTo interactive notebook.
We illustrate the main functionalities of GINsim on a reference multilevel model of regulatory network that control the mammalian cell cycle, including the computation of its stable states and cyclic attractor, the assessment of their reachability, and the identification of the regulatory circuits driving multistability and oscillations.
GINsim is freely available under GPL-3.0 at \url{https://ginsim.github.io/}, runs on any platform supporting Java, and is interoperable with the other tools of the CoLoMoTo community.
Its website provides installation instructions and reference documentation, as well as a repository of Boolean and multilevel logical models.
}



\maketitle

\section{Introduction}\label{Introduction}

Boolean modelling approaches are becoming increasingly popular for integrating our current knowledge about large and complex cellular signalling and regulatory networks in order to explore their dynamical behaviours (see, e.g., \cite{AbouJaoude2016, Zanudo2018, Hemedan2022}, and references therein). Over the last decades, many software tools have been developed to ease the delineation and the analysis of comprehensive Boolean models (see, e.g., \cite{Noel2025colomoto}), or to infer such models directly from high-throughput molecular data (for a recent review, see \cite{Klemmer2026}).
In parallel, many models of signalling and regulatory networks governing cell fate decisions have been published and are currently available in dedicated repositories, such as CellCollective ({\href{https://research.cellcollective.org/research/dashboard/}{https://research.cellcollective.org/research/dashboard/}}), Biodivine ({\href{https://github.com/sybila/biodivine-boolean-models}{https://github.com/sybila/biodivine-boolean-models}}), or GINsim's collection ({\href{https://ginsim.github.io/models/}{https://ginsim.github.io/models/}}).

Developed in Java, the software GINsim (for Gene Interaction Network simulation) has been maintained and updated regularly over the past twenty years, with the addition of various novel functionalities \cite{Gonzalez2006ginsim,Naldi2009ginsim,Chaouiya2012ginsim,Naldi2018ginsim}.
GINsim version 3.1 is based on the multilevel formalism initially proposed by René Thomas \cite{Thomas1991,Chaouiya2003}. It runs on any major operating system for which a Java Virtual Machine (version 11 or later) is available. It is distributed as a single Java ARchive (JAR) file, freely available from the Installation section of the GINsim website (\url{https://ginsim.github.io/install/}). Once installed, GINsim can be launched by double-clicking on the file or from a terminal with the command \texttt{java -jar GINsim-\#version.jar}.

Hereafter, we first review some of the main functionalities of GINsim 3.1, focusing in particular on those enabling the dynamical analysis of large logical models. Next, we showcase selected functionalities by applying them to a model of the core engine controlling the mammalian cell cycle, as defined by Traynard et al \citep{Traynard2016}.
Readers interested in delineating novel models from scratch are referred to the online GINsim tutorial (\url{https://ginsim.github.io/tutorial/}, see also \cite{Naldi2018ginsim}).

GINsim's architecture is organised in three layers:
MDDlib, a low-level library for manipulating logical functions using multi-valued decision diagrams\footnote{\url{https://github.com/colomoto/mddlib}};
bioLQM \cite{Naldi2018biolqm}, a Java library that builds on MDDlib to provide model manipulation, format conversion, and the core analysis algorithms, shared with other tools (e.g. EpiLog \cite{Varela2019epilog});
and the GINsim graphical user interface (GUI) itself, which builds on bioLQM to provide interactive model editing, visualisation, and analysis.
This layered modular design allows the core algorithms to be reused or combined outside of the GINsim GUI.
For example, from the command line, programmatically as a Java library, or from Python/Jupyter notebooks through the CoLoMoTo environment, which offers an integrated access to over twenty Boolean modelling tools \cite{CoLoMoTo2015,Naldi2018notebook, Noel2025colomoto}.

\begin{table}[t]
\resizebox{\textwidth}{!}{%
\begin{tabular}{c|l|l|c|l}
    \textbf{Menu}& \textbf{Functionality} & \textbf{Description} & \textbf{Graph type} & \textbf{Status} \\ \hline
    \multirow{10}{*}[-3em]{\rotatebox{90}{Tools}} & Run simulation & Computation of transition graphs (STG, SCCG, HTG) & LRG & Maintained \\ \cmidrule{2-5}
    & Attractors reachability & Monte Carlo and Avatar algorithms for reachability probabilities & LRG & Extended \\ \cmidrule{2-5}
    & Compute stable states & Compute the logical model stable states & LRG & Extended \\ \cmidrule{2-5}
    & Trap-spaces identification & Compute stable patterns in the state space & LRG & New \\ \cmidrule{2-5}
    & Analyse circuits & Compute circuit functionality & LRG & Maintained \\ \cmidrule{2-5}
    & Interactions functionality & Recompute the sign of interactions from logical rules & LRG & Maintained \\ \cmidrule{2-5}
    & Reduce model & Perform several model reductions (outputs or specific nodes) & LRG & Extended \\ \cmidrule{2-5}
    & Booleanise model & Create an equivalent Boolean model & LRG & New \\ \cmidrule{2-5}
    & Reverse model & Create a model capable of generating the reverse dynamics & LRG & New \\ \cmidrule{2-5}
    & Expand SCC\&HTG nodes & Recover the original graph features from a selection of SCC or HTG graph nodes & SCCG/HTG & Maintained \\ \hline
    \multirow{5}{*}[-1em]{\rotatebox{90}{Graph}}
    & Find path & Find a path between two or more nodes/states in the graph & LRG/STG/SCCG/HTG & Extended \\ \cmidrule{2-5}    
    & Colour LRG & Colour a graph depending on the state of each regulator & LRG & Extended \\ \cmidrule{2-5}
    & Colour SCC & Identify and colour SCC in the graph & LRG/STG/SCCG/HTG & Maintained \\ \cmidrule{2-5}
    & Construct SCC graph & Construct the SCC graph from the original graph & LRG/STG/SCCG/HTG & Maintained \\ \cmidrule{2-5}
    & Animate & Animate the regulatory graph following a path in the STG & STG & Maintained \\ \hline
\end{tabular}}
\caption{List of GINsim functionalities contextually available for the different types of graphs supported: logical regulatory graphs (LRG), state transition graphs (STG), strongly connected component graphs (SCCG), and hierarchical transition graphs (HTG).
Status with respect to the features described in previous GINsim articles \cite{Gonzalez2006ginsim,Naldi2009ginsim,Chaouiya2012ginsim,Naldi2018ginsim}: \emph{New} (not described before), \emph{Extended} (described before, with new options or algorithms), \emph{Maintained} (described before, unchanged).}
\end{table}

Upon GINsim launch, the user can create a new model or open a previously saved model (in \texttt{.ginml} or \texttt{.zginml} formats), or import a model from a supported file format (\texttt{.sbml}, \texttt{.bnet}, \texttt{.tt}, \texttt{.net}, \texttt{.cnet}).

\begin{figure}[!t]
    \centering
    \includegraphics[width=0.8\linewidth]{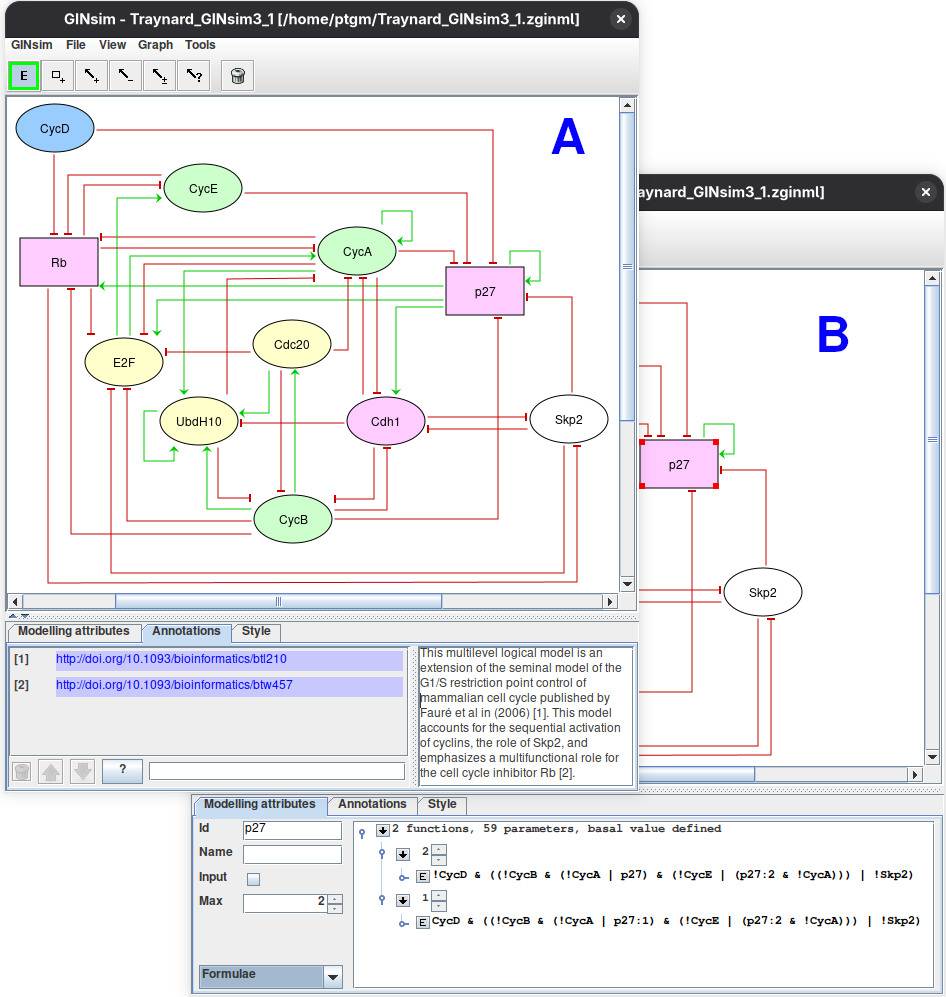}
    \caption{\textbf{Graphical interface of GINsim.}
\textbf{A}: The top of the GUI encompasses several menus providing access to the different functionalities of GINsim, while graph editing buttons are found just below.
The main window contains the regulatory graph for the mammalian cell cycle engine proposed by Traynard et al \cite{Traynard2016}, while the model attributes (name, list of nodes) and annotations (bibliographic references and free text) are displayed below.
Rectangles denote the ternary nodes (Rb and p27) and ellipses denote Boolean nodes. The input node CycD is shown in blue. Green and red arcs denote activations and inhibitions, respectively.
\textbf{B:} Two (non-overlapping) Boolean rules are associated with the ternary node p27, as displayed below the regulatory graph. The first rule enables the value 2, while the second one enables the value 1. These rules combine the identifiers of the regulators of p27 with the classical operators NOT (!), AND (\&), and OR ($\mid$). By default, all unspecified regulatory input situations result in the assignment of the value 0.}
    \label{figure1}
\end{figure}

The user can edit the logical regulatory graph by adding/removing nodes and edges.
Clicking on a node or an edge, the bottom area enables changes in the node/edge characteristics through three tabs: \emph{Modelling attributes} (whose content depends on the selected element), \emph{Annotations} (to annotate the model, its nodes and its edges) and \emph{Style} (governing appearance) (see Figure~\ref{figure1}).

\section{Methods}\label{sec:methods}

\paragraph{Logical regulatory graph (LRG)}
In GINsim, Boolean or multilevel models are implemented in terms of directed, signed graphs, where vertices (also called nodes of the LRG) denote regulatory proteins or genes, while directed, signed arcs denote regulatory interactions (inhibitions, activations, or context-sensitive interactions, which can be labelled as dual or as unknown). By default, the activity of the regulatory nodes is modelled by Boolean variables, taking the value $0$ when the node is considered inactive, and the value $1$ when the node is considered active.
However, multilevel variables can be used when biologically justified. Boolean rules are then associated with each node to specify the regulatory requirement to reach a given activity level. These rules combine literals corresponding to the upstream regulatory nodes with the logical operators \emph{NOT} (!), \emph{AND} (\&) and \emph{OR} ($\mid$).

GINsim GUI supports the delineation of such regulatory graphs from scratch, including the logical rules and annotations of the model with free text and hyperlinks to public database entries. However, since this article focuses mainly on novel functionalities enabling the dynamical analysis of large models, we refer interested readers to GINsim documentation to learn how to delineate a LRG (see also \cite{Naldi2018ginsim}). 

\paragraph{Annotations}
To document a model and keep track of the knowledge supporting it, annotations can be attached to the model as a whole (by selecting the background of the canvas), to any node, or to any interaction. The annotation of the selected element is displayed in the \emph{Annotations} tab, and combines a numbered list of links to external resources with a free text, in which numbered links can be cited.
Links are entered as full web addresses (e.g., \url{https://doi.org/10.3389/fphys.2018.00646}) or as compact identifiers using the prefixes of the \url{https://identifiers.org} registry~\cite{Juty2012}, such as \texttt{pubmed:29971008} or \texttt{doi:10.3389/fphys.2018.00646} for publications, or \texttt{hgnc:11998} for a gene (here TP53); clicking on a compact identifier opens the corresponding identifiers.org address, which redirects to the relevant website. Annotations are stored with the model in the \texttt{.zginml} archive.

\paragraph{Perturbations}
Experimental or hypothetical perturbations are defined at the level of the logical model and stored with it in the \texttt{.zginml} archive. They are implemented as model modifiers in the bioLQM library \cite{Naldi2018biolqm}, and can thus also be applied programmatically.
A simple perturbation locks the value of a node to a given level, or to a range of levels: a knockout (loss of function) locks it to 0, whereas an ectopic expression (gain of function) locks it to 1 or above for a multilevel node.
More subtle perturbations modify the rule of a target node to make it insensitive to the activity of a given regulator (e.g. E2F[Rb@0] in Table~\ref{tab:perturbations}).
Simple perturbations can also be combined into multiple perturbations.
A defined perturbation can then be selected in most of the analyses described below (simulation, stable states, trap-spaces, attractors reachability, interaction functionality, and  circuit analysis), which are performed on the perturbed model, with the original model left unchanged.

\paragraph{Computation of a state transition graph}
The dynamics of a Boolean model is often represented in terms of a state transition graph (STG), whose vertices correspond to Boolean states (i.e., vectors containing specific values for all regulatory components), and arcs correspond to transitions between pairs of states, enabled by the Boolean rules for a given updating scheme (most often fully asynchronous or synchronous).
After defining or loading a logical regulatory graph (LRG) in GINsim, the user can analyse its dynamics by running a simulation, via the \emph{Run simulation} entry of the \emph{Tools} menu. This computes the STG, or a compression thereof, representing all the reachable states of the model and their admissible transitions, under a chosen updating mode (e.g., synchronous, asynchronous, or based on user-defined priority classes), optional perturbations, and selected initial states. Noteworthy, GINsim further supports the use of a \emph{complete} updating mode, which considers all combinations of updating calls in each state.

\paragraph{Computation of stable states}
Logical stable states typically represent persistent cellular types or phenotypes and are thus of particular interest to modellers.
In this respect, GINsim implements a symbolic, simulation-free procedure that enables the exhaustive identification of all stable states (fixed points) of a logical model, based on decision diagrams \cite{Naldi2007CMSB}. Available from the \emph{Tools} menu, the \emph{Compute stable states} functionality can be applied with the optional selection of a user-defined node perturbation and/or predefined input restrictions, and the results are then displayed in a dedicated table. This functionality is delegated to the bioLQM library \cite{Naldi2018biolqm}, allowing its programmatic use outside the graphical interface.

\paragraph{Identification of trap-spaces}
Trap-spaces are subspaces of the state space that cannot be escaped by the model dynamics, independently of the updating mode; (minimal) trap-spaces provide tractable over-approximations of the attractors of a model, encompassing both stable states and complex (cyclic) attractors. GINsim computes trap-spaces symbolically, without simulation, through the \emph{Trap-spaces identification} option of the \emph{Tools} menu. Here also, the user can optionally select a predefined perturbation or input restriction. Based on the algorithm proposed by Klarner and collaborators \cite{Klarner2014}, this feature is implemented through the bioLQM back-end.

\paragraph{Attractors reachability}
GINsim provides a dedicated tool to estimate the reachability probabilities of attractors (stable states and complex attractors) from selected sets of initial states, without requiring the exhaustive construction of the state transition graph.
This analysis relies on sampling of asynchronous trajectories, either by the Monte Carlo approach or by using the Avatar algorithm \cite{Mendes2018avatar}, which is better suited to deal with complex cyclic behaviour.
This analysis is accessible from the \emph{Tools} menu and can be combined with perturbations to explore how mutations affect the reachability of attractors.

In contrast to MaBoSS \cite{Stoll2017maboss}, which performs continuous-time stochastic simulations of Boolean models and computes the temporal evolution of state distributions, the Avatar algorithm focuses on the probability of reaching each attractor.
The Avatar algorithm explicitly handles the trajectories trapped in large cyclic structures, which would otherwise render the simple simulation inefficient. Moreover, Avatar naturally supports Boolean and multi-valued models alike.

\paragraph{Interactions functionality}
The delineation of a logical model with GINsim starts with that of a directed, signed regulatory graph, which is later complemented by the definition of Boolean rules, internally encoded as decision diagrams.
When defining a node rule, GINsim does verify that all literals included correspond to source regulatory nodes and raises an error when it is not the case. 
However, this verification mechanism does not preclude discrepancies between the interaction signs and the rules.
In order to help the user identify potential inconsistencies, we have implemented the \emph{Interactions functionality} tool.
Available in the \emph{Tools} menu, this functionality computes the signs of each interaction for each regulatory context directly from the Boolean rules, either for selected nodes or for the whole model.
The recomputed signs can then be saved in an HTML file, and discrepancies can be further displayed directly on the regulatory graph (for more details, see the Documentation of GINsim).
This functionality is particularly useful for the delineation of consistent rules for multilevel nodes, as well as for Boolean nodes with complex rules.

\paragraph{Analyse circuits}
Regulatory circuits are defined as simple circular sequences of regulatory interactions. Depending on the parity of the number of negative interactions involved, regulatory circuits are classified as negative (involving an odd number of negative interactions) or positive (involving an even number of negative interactions) circuits. Positive circuits are a prerequisite for multistability, whereas negative circuits underlie periodic behaviour \cite{Thieffry2007,Remy2008}. However, the dynamical role of such circuits depends on the rules associated with the nodes involved.
In this respect, GINsim includes an original tool to identify all or subsets of circuits in a model, determine which of these circuits are dynamically functional (i.e., contribute to the generation of multistationary or periodic behaviour) and further delineate the corresponding \emph{functionality context} in terms of conditions on the values of nodes regulating the nodes involved in the circuit \cite{Naldi2007CMSB}. This analysis helps identify which of the (often numerous) circuits of a regulatory graph actually contribute to its dynamics. Available from the \emph{Analyse circuits} entry of the \emph{Tools} menu, this functionality allows the user to restrict the search to circuits satisfying length or component constraints and to repeat the analysis for predefined genetic perturbations.

\paragraph{Model reduction}
The size of STG increases exponentially with that of LRG, impeding the dynamical analysis of large and complex networks. To avoid this issue, one approach consists in reducing the model size. In this respect, the \emph{Reduce model} entry of the \emph{Tools} menu compresses a regulatory graph by hiding a chosen subset of components, while the logical rules of their targets are updated to account for the indirect effects of the hidden regulators. This construction preserves the most relevant dynamical properties of the original model, including its stable states and more complex attractors. The correspondence between the attractors of the original and reduced models is formally established \cite{Naldi2011reduction}. This reduction functionality further supports automatic stripping of output components (i.e., components without outgoing regulatory arcs), which do not affect the set of attractors.

\paragraph{Model Booleanisation}
Several useful algorithms are currently limited to the analysis of Boolean models (e.g., the trap-space identification algorithm mentioned above). However, multi-valued logical models can be converted into equivalent Boolean models, using the mapping proposed by van Ham \cite{vanHam1979}, in which each component with maximal level $m$ is encoded by $m$ Boolean variables. The resulting Boolean model reproduces the dynamical behaviour of the original multi-valued model, while additional safeguards prevent the exploration of non-admissible states and the introduction of spurious non-admissible attractors. Implemented in bioLQM, this Booleanisation can be accessed in GINsim through the \emph{Tools} menu, generating an equivalent Boolean model, or implicitly whenever a multi-valued model is submitted to a Boolean-only algorithm (such as the identification of trap-spaces) or exported to a Boolean-only format.

\paragraph{Model reversal}
Once an attractor has been identified, one is often interested in computing the ensemble of states leading to that attractor. This can be best achieved by tracing the transitions and the corresponding states leading to this attractor. In this respect, GINsim includes a \emph{Reverse model} tool enabling the generation of a model whose asynchronous successors correspond to the predecessors of the original model, effectively reversing the direction of the dynamics. Multi-valued models are handled by first booleanising them, with additional transformations to avoid the generation of non-admissible successor states. This feature is implemented in bioLQM and is accessible from the \emph{Tools} menu.

\paragraph{Find path}
The \emph{Find path} option of the \emph{Graph} menu computes the paths between two or more nodes of the active LRG, or between two or more nodes of the active STG, SCCG or HTG. The paths found are highlighted in the corresponding graph, while the corresponding sequence of nodes is listed in the interface.

\paragraph{Animate}
The STG animator enables the interactive, graphical exploration of a trajectory (path) within an existing state transition graph by manually selecting successive transitions from a given state. It is available from the \emph{Graph} menu when a state transition graph is active.

\paragraph{Colour LRG}
This tool highlights the components whose values are consistent with a given reference state or pattern on the regulatory graph, thereby providing a visual means for inspecting a state (e.g., a stable state or an initial condition) directly on the network structure. It is available from the \emph{Graph} menu.

\paragraph{Colour SCC}
Directed graphs can be compressed by grouping the vertices (i.e., model nodes or states) with the same reachability properties. A strongly connected component (SCC) is a maximal subgraph in which every node is reachable from all other nodes.
Available from the \emph{Graph} menu, the \emph{Colour SCC} functionality  colours the components of a graph according to the strongly connected component (SCC) they belong to, offering an immediate visual overview of the cyclic structure of the graph prior to (or instead of) an explicit computation of the SCC graph.

\paragraph{Compression of regulatory and state transition graphs}
Based on the notion of SCC, any directed graph can be compressed into  an acyclic graph obtained by contracting each SCC of the original graph into a single node. The resulting strongly connected component graph (SCCG) thereby provides a compact overview of the organisation of the original graph.
GINsim includes a tool for the computation of the SCCG that can be applied to all kinds of supported graphs (i.e., regulatory graphs, state transition graphs, and compressions thereof).
Furthermore, GINsim implements a stronger graph compression into  \emph{hierarchical transition graph} (HTG), which further groups the vertices leading to identical sets of strongly connected components of the corresponding SCCG \cite{Berenguier2013}. 
Based on the classical Tarjan algorithm \cite{Tarjan1972},  SCCG and HTG computations are typically applied to state transition graphs to delineate attractors. The computation of a SCCG can be launched with the \emph{Construct SCC graph} option in the \emph{Graph} menu, while the compression of  a STG into a SCCG or a HTG can be specified by selecting the corresponding entry of the \emph{Construction strategy} rolling menu in the \emph{Run simulation} interface. 
Conversely, the \emph{Expand SCC\&HTG nodes} action of the \emph{Tools} menu allows the recovery of the original graph features from a selection of SCC or HTG graph nodes, at the cost of launching a new simulation limited to the corresponding subspace.

\paragraph{Scripting and integration in the CoLoMoTo notebook}
Besides the graphical interface, GINsim can be driven by scripts.
The \texttt{-s} option of the launcher runs scripts, for instance written in Jython, which can load models, call the GINsim services (e.g., simulation, stable states, SCC graph, interaction analysis, image export) and use the tools and modifiers of the bioLQM library, such as the perturbations.
In addition, the \texttt{-py} option starts GINsim as a Python gateway (based on Py4J), which is used by the \texttt{ginsim} Python module of the CoLoMoTo interactive notebook \cite{Naldi2018notebook,Noel2025colomoto}.
From a Jupyter notebook, a model can thus be loaded, displayed, perturbed, analysed, and converted into the formats of other tools of the environment (e.g., MaBoSS, Pint or pyboolnet), within a single reproducible workflow.
For example, the stable states and the trap-spaces of Figure~\ref{figure2} can be obtained as follows:
\begin{verbatim}
import ginsim, biolqm
lrg = ginsim.load("Traynard_MultiLevel_MamCC_Apr2016.zginml")
lqm = ginsim.to_biolqm(lrg)    # converts the model in the bioLQM format
biolqm.fixpoints(lqm)          # computes the two stable states
pert = biolqm.perturbation(lqm, "Rb%0 E2F%1")
biolqm.trapspace(pert, "bdd")  # identifies the two cyclic trap-spaces
\end{verbatim}
Notebook-based tutorial on GINsim is available on the CoLoMoTo Docker website: \url{https://colomoto.github.io/colomoto-docker/tools/ginsim.html}.

\paragraph{Online model repository}
GINsim is distributed with a model repository (\url{https://ginsim.github.io/models/}), which currently gathers around seventy published logical models spanning three decades of work, from the seminal phage $\lambda$ lysis/lysogeny switch model \cite{Thieffry1995} to recent models of cancer signalling and stem cell differentiation. The repository can be browsed through a searchable table indexing each entry by taxon (bacteria, yeast, \emph{Drosophila}, \emph{C. elegans}, zebrafish, sea urchin, mammals/human, \ldots), by biological process (cell cycle, cell differentiation, cell-fate decision, development, signalling, cancer, senescence, sex determination, etc), by publication year, and by author(s). 
Each entry provides a link to a dedicated page including bibliographic references, a summary of the model and its main results, illustrative figures (regulatory graph, state transition graphs, attractors) and the corresponding model file(s), ready to be downloaded and opened with GINsim.
This repository serves both as a curated collection of reference case studies to benchmark new tools and as a resource for the community to retrieve, reuse and extend published models.

\section{Case study}\label{sec:case-study}
Following the seminal modelling studies by Novak, Tyson \emph{et al.} (see, e.g. \cite{Chen2000,Novak2001,Novak2004}), it turned out to be possible to reproduce the qualitative dynamical properties of systems of non-linear ordinary differential equations using the simpler Boolean framework \cite{Faure2006,Davidich2008,Li2004,Irons2009}, or moderate multilevel extensions thereof \cite{Faure2009, Mombach2014, Traynard2016}. 
Hereafter, we use the multilevel model defined by Traynard et al \cite{Traynard2016} to demonstrate some of the key functionalities of GINsim, including the computation of model attractors and their visualisation, the construction and compression of dynamical graphs, the analysis of attractor reachability, and the identification of functional circuits, all of these potentially considering initial conditions, as well as simple or multiple node perturbations. 
This cell cycle model is available in the GINsim model repository at \url{https://ginsim.github.io/models/2016-mammal-cell-cycle/}.

\paragraph{Multilevel logical model for the mammalian cell cycle}
Shown in \textbf{Figure 1A}, the model published by Traynard \emph{et al.} encompasses eleven components involved in the control of the proliferation of mammalian cells, including cyclins A, B, D and E. 
In brief, Cyclin D (CycD, an input in this model) is the main target of the growth factors that drive cells out of the quiescent state to enter the cell cycle. Cyclin E (CycE) regulates the transition between the first gap phase and the DNA synthesis (S) phase. Cyclin A (CycA) controls the S phase and its progression into the second gap phase. Finally, Cyclin B (CycB) triggers the transition from the G2 phase to mitosis, i.e., cell division \textit{per se}. 
The model also encompasses three key inhibitors of the cell cycle: the retinoblastoma protein Rb, the Cdk inhibitor p27/Kip1 (p27) and the proteasome complex represented by its co-activators Cdh1 and Cdc20.
The model further involves the E2 ubiquitin conjugating enzyme UbcH10, which participates in Cdh1-dependent degradation of cyclin A.

The main documented interactions between these components can be visualised in the LRG shown in Figure~\ref{figure1}. For appropriate logical rules (as defined in Figure~\ref{figure1}-B and Supplementary Table~\ref{tab:logical_rules}), this model accounts for a cyclic attractor characterised by the periodic activities of Cyclins E, A and B, which drive the cells through key transitions by enabling the phosphorylation of a number of substrates by their catalytic partners, the cyclin-dependent kinases (CDKs, implicit in this model). 

Interestingly, the model further accounts for the impact of various documented genetic perturbations of the cell cycle engine. From the logical modelling point of view, knocking-out a gene simply amounts to locking the value of the corresponding node to 0, while forcing the ectopic activation of a node amounts to locking its value to 1. GINsim supports the definition and storage of such perturbations along with the logical model and further eases their combinations to define multiple perturbations. A list of documented perturbations relevant for the mammalian cell cycle model with their associated phenotypes can be found in Table~\ref{tab:perturbations}.

\begin{table}
\resizebox{\textwidth}{!}{
\begin{tabular}{l|p{8.5cm}|c}
\hline
\textbf{Perturbations} & \textbf{Phenotypes} & \textbf{References}\\ \hline
Rb KO & Cycling in absence of growth factor & \cite{Faure2006} \\ \hline
Cdh1 KO & Cycling in presence or absence of growth factors & \cite{Faure2006} \\ \hline
p27 ectopic expression & Cell cycle arrest in presence of growth factors & \cite{Faure2006} \\ \hline
E2F ectopic in Rb$^{-/-}$ & Cyclic behaviour in presence or absence of growth factors & \cite{Faure2006} \\ \hline
E2F ectopic in Rb$^{+/+}$ & Targets of E2F are expressed in absence of growth factor but cell cycle is then arrested & \cite{Traynard2016} \\ \hline
UbcH10 KO & Cell cycle arrested in M phase, destruction of cyclin A and B inhibited & \cite{Traynard2016} \\ \hline
p27 KO & Viable, less serum-dependent & \cite{Traynard2016} \\ \hline
CycE KO & Cell cycle arrest before S phase & \cite{Traynard2016} \\ \hline
RbR661W (E2F[Rb@0]) & Cell cycle in presence of growth factors, cell cycle arrest in absence & \cite{Traynard2016} \\ \hline
Skp2 KO & Cell cycle arrest or endoreplication with Cyclin E and p27 accumulation & \cite{Traynard2016} \\ \hline
Skp2/p27 double KO & Cell cycle in presence of growth factors, cycle arrest in their absence & \cite{Traynard2016} \\ \hline
\end{tabular}}
    \caption{Documented perturbations of the mammalian cell cycle and associated phenotypes. For a knockout (KO), the value of the corresponding node is locked to 0; while for an ectopic expression, the value of the corresponding node is locked to 1 (or above for multi-valued nodes). E2F[Rb@0] denotes a modification of the E2F rule making E2F insensitive to Rb activity. The reported phenotypes are indicated together with citations of the logical models qualitatively recapitulating them.}
    \label{tab:perturbations}
\end{table}

\begin{figure}[!t]
    \centering
    \includegraphics[width=1\linewidth]{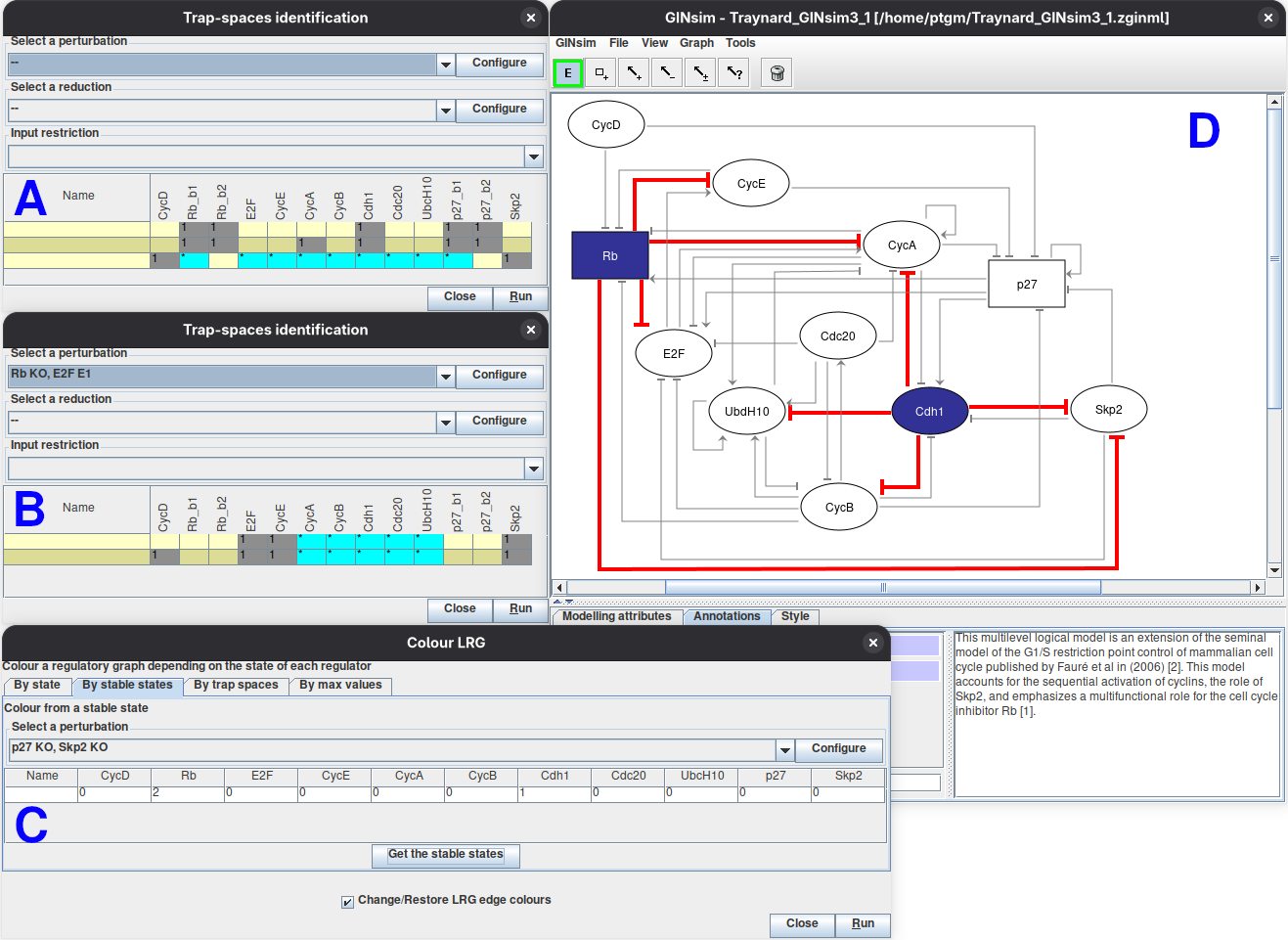}
    \caption{\textbf{Computation and visualisation of the trap-spaces for the mammalian cell cycle model.}
\textbf{A}: In the absence of perturbation, running the \emph{Trap-spaces identification} functionality (available from the \emph{Tools} menu) leads to two stable states when the input node CycD is inactive, differing only in the value of CycA, and to one cyclic attractor when CycD is active, with nine oscillating nodes. Values 0, 1 and free are denoted by empty yellow cells, grey cells with the value \emph{1} and blue cells with a \emph{*}, respectively. Note that the two ternary components Rb and p27 are each represented by a pair of Boolean variables (Rb\_b1, Rb\_b2 and p27\_b1, p27\_b2), as the trap-space identification algorithm is currently restricted to Boolean models.
\textbf{B:} In the presence of a double perturbation corresponding to a knockout of Rb and an ectopic activation of E2F (\emph{Rb KO, E2F E1}), the computation of trap-spaces leads to two cyclic attractors, differing in the value of CycD, which implies that the model predicts cyclin oscillations even in the absence of growth factors, as observed experimentally.
\textbf{C: }Interface of the \emph{Colour LRG} functionality (available from the \emph{Graph} menu) with the selection of a stable state to be displayed on the regulatory graph (tab \emph{By stable states}) in the context of a double knockout affecting p27 and Skp2. The stable state returned is characterised by the values Rb = 2 and Cdh1 = 1, all the other components being inactive.
\textbf{D}: Result of the execution of the \emph{Colour LRG} functionality as specified in panel C, denoting the active nodes (Rb and Cdh1) and inactive nodes with dark blue and white backgrounds, respectively. The interactions exerted by the two active nodes (all inhibitory in this case) are highlighted in red.}
    \label{figure2}
\end{figure}

\paragraph{Computation and visualisation of the attractors of the model}
As mentioned in the Methods section, GINsim 3.1 implements an efficient symbolic method for the computation of trap-spaces \cite{Klarner2014}. 
Figure~\ref{figure2}-A displays the results returned after running the \emph{Trap-spaces identification} tool for the mammalian cell cycle model with default parameters (i.e., in the absence of any model perturbation, reduction, or input restriction). The first row of the table corresponds to a stable state with all cell cycle inhibitors activated and all cyclins inactivated, accounting for cell quiescence. The second row corresponds to a very similar stable state, but with CycA activated. This spurious stable state tentatively points to incomplete information on the regulation of Cyclin A \cite{Traynard2016}. Finally, the third row corresponds to a complex cyclic attractor, in the presence of CycD, with Skp2 activated and all other components oscillating between the values 0 and 1.
In contrast, Figure~\ref{figure2}-B displays the trap-spaces obtained when selecting a perturbation that locks Rb to the value 0 and E2F to the value 1. This double perturbation annihilates the stable states and leads to two cyclic attractors, one for each value of the input CycD, as observed experimentally.
Figure~\ref{figure2}-C displays the graphical interface obtained after selecting the \emph{Colour LRG} functionality from the \emph{Graph} menu, selecting the double p27/Skp2 KO, clicking successively on the \emph{By stable states} tab and on the \emph{Get the stable states} button, and finally selecting the stable state returned. Following these actions, the regulatory graph is coloured (Figure~\ref{figure2}-D) to emphasise the active components (Rb and Cdh1, in blue) and the active interactions (here, all inhibitory interactions shown in red). Such a visual representation is particularly useful when dealing with large models.

\begin{figure}[!t]
    \centering
    \includegraphics[width=1\linewidth]{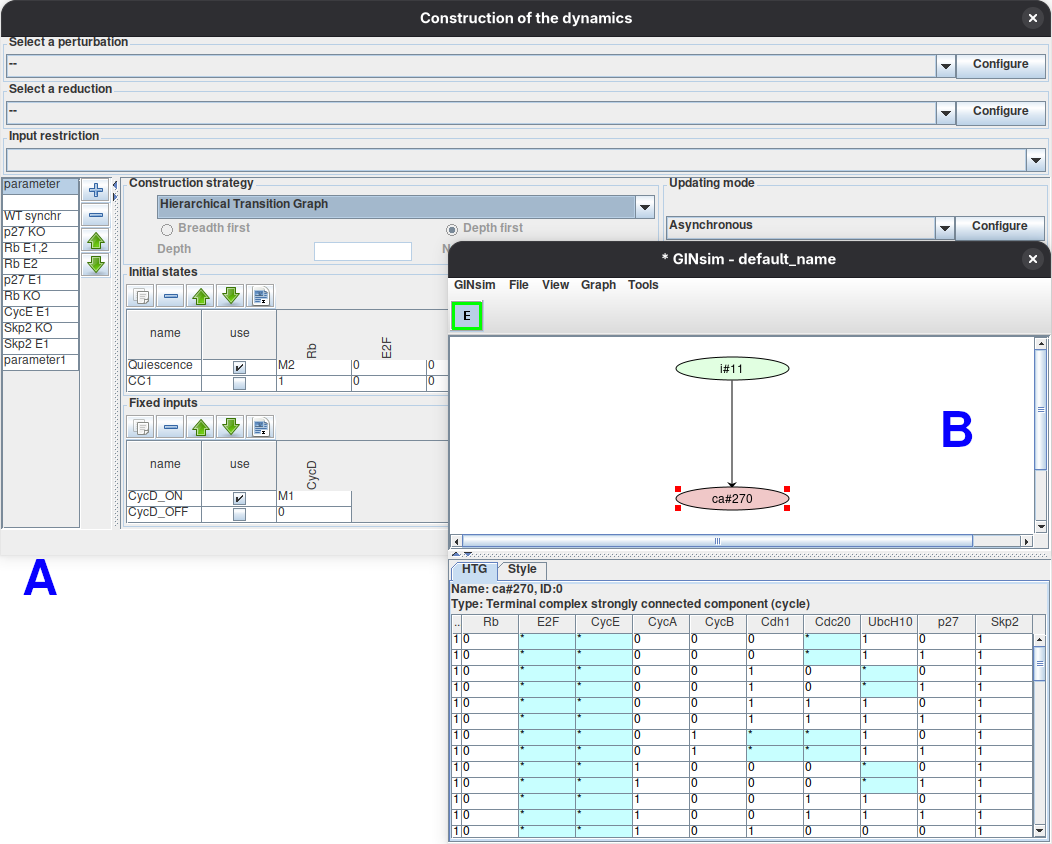}
    \caption{\textbf{Construction of state transition graphs or compression thereof.} 
\textbf{A}: Interface enabling the configuration and the computation of the dynamical behaviour of the mammalian cell cycle model (activated with the \emph{Run simulation} functionality available from the \emph{Tools} menu), with the possibility to select (and eventually define) a perturbation, a model reduction, the type of graph to be computed (here a \emph{Hierarchical Transition Graph}), the updating mode (here \emph{Asynchronous}), and the delineation and selection of an initial state. Here, the initial state \emph{Quiescence} is combined with the input configuration CycD\_ON, which corresponds to a quiescent situation with CycD active. Note that several configurations can be stored with user-defined names in the \emph{parameter} area (left).
\textbf{B}: Hierarchical transition graph (HTG) obtained by running a simulation with the configuration displayed in panel A. The top component of this graph contains eleven transient states (\emph{i} stands for irreversible, while \emph{\#11} denotes the size of the state set), which ultimately lead to the bottom component of the HTG, itself containing 270 states forming a complex cyclic attractor (\emph{ca} stands for cyclic attractor, while \emph{\#270} denotes the size of that attracting state set). The states belonging to the cyclic attractor are displayed below, in terms of a series of patterns, where the \texttt{*}'s denote free variables.}
    \label{figure3}
\end{figure}

\paragraph{Construction and compression of state transition graphs}
Figure~\ref{figure3}-A displays the interface obtained after selecting the \emph{Run simulation} functionality from the \emph{Tools} menu. This interface supports the delineation of user-defined initial states for the internal nodes of the network, as well as of combinations of values for the input nodes. Here, an initial state is selected, corresponding to a quiescent state with CycD activated. The user can further select a predefined model perturbation (or launch the window enabling the delineation of perturbations), a model reduction, and an updating mode (cf. Methods). Finally, the user can choose between three construction strategies to generate a STG, a SCCG, or a HTG, the latter being selected here. To ease the exploration of the dynamics of large models, it is further possible to choose between a depth-first and a breadth-first exploration and limit the depth of paths or the number of states to visit.

The launch of a simulation of the cell cycle model with the configuration shown in Figure~\ref{figure3}-A results in the compressed \emph{hierarchical state transition graph} (HTG) shown in Figure~\ref{figure3}-B. The upper graph node groups eleven states (including the initial state) all ultimately leading to the lower node, which groups 270 states forming a complex cyclic attractor, tentatively over-approximating the periodic behaviour of wild-type mammalian cells. The states belonging to this asynchronous cyclic attractor are listed in the panel just below the HTG, in terms of a series of patterns.
The \emph{Expand SCC\&HTG nodes} functionality available from the contextual \emph{Tools} menu further enables the expansion of selected HTG (or SCCG) nodes into the corresponding state transition graph.

\begin{figure}[!t]
    \centering
    \includegraphics[width=1\linewidth]{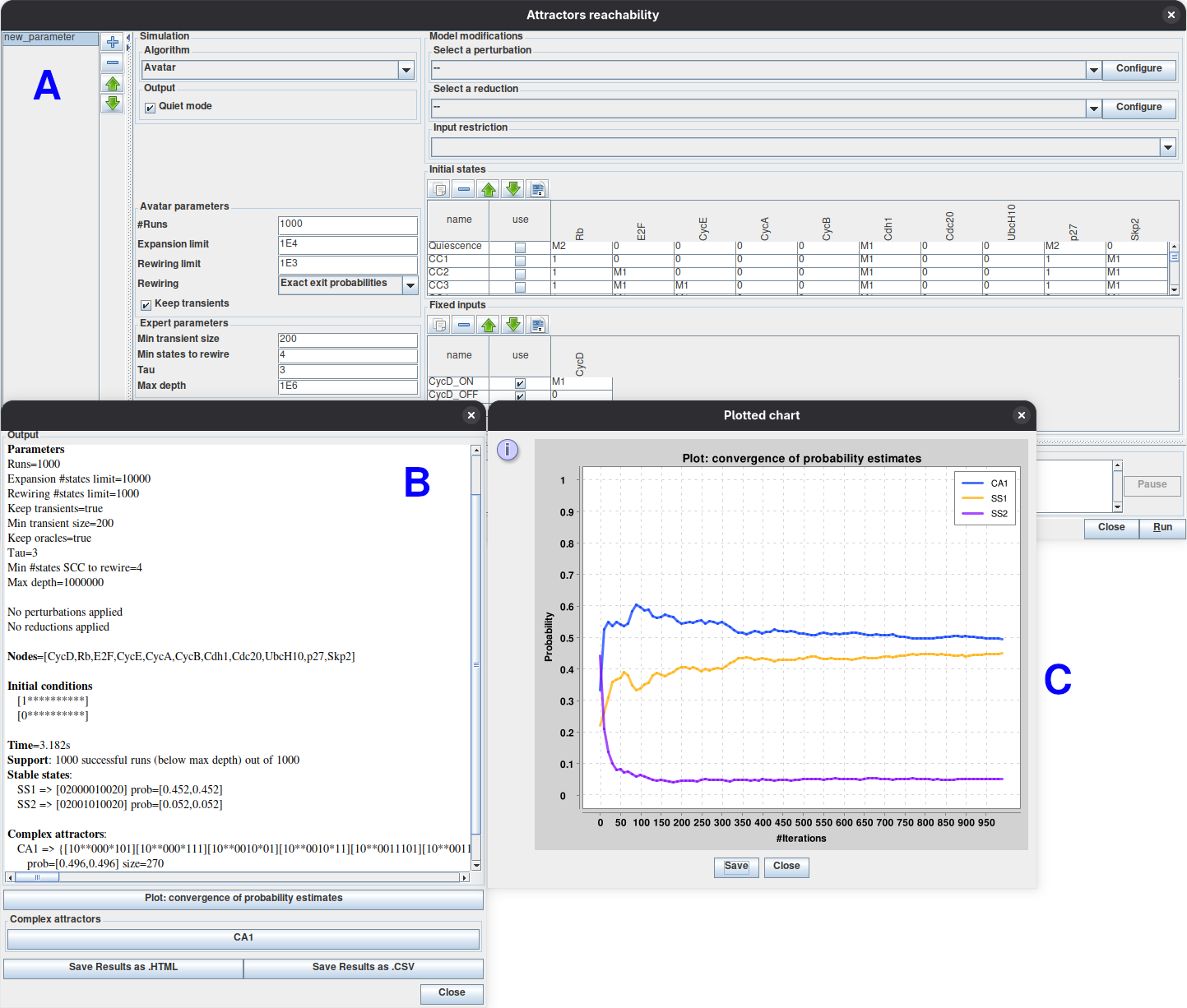}
    \caption{\textbf{Estimation of the reachability probabilities of attractors.}
\textbf{A}: Interface of the \emph{Attractors reachability} tool (available from the \emph{Tools} menu), enabling the selection of the Avatar algorithm and of its parameters (here 1000 runs, with limits on the number of states expanded and rewired, and rewiring based on exact exit probabilities). This interface further enables the selection of model modifications (perturbation, reduction, input restriction; none considered here), as well as of initial conditions (here no constraint on the internal or input components).
\textbf{B}: Output window reporting the parameters used, the initial conditions (where \emph{*} denotes free variables) and the estimated reachability probabilities of the attractors: the two stable states SS1 and SS2 (with reaching probabilities equal to 0.452 and 0.052, respectively; first two rows of Figure~\ref{figure2}-A) and the complex cyclic attractor CA1 (270 states, probability of 0.496), based on 1000 successful runs. The results can be saved as HTML or CSV files.
\textbf{C}: Convergence of the probability estimates for CA1, SS1 and SS2 as a function of the number of iterations.}
    \label{figure4}
\end{figure}

\paragraph{Delineation of the reachability of the attractors}
Beyond identifying the attractors themselves, it is often important to know whether a model can reach a given attractor from a given initial condition, with some probability estimation, and how this is affected by perturbations.
Figure~\ref{figure4} illustrates the use of the \emph{Attractors reachability} tool (available in the \emph{Tools} menu) on the mammalian cell cycle model.
Starting from initial conditions in which all components (internal and input) are left unconstrained (their values are sampled for each trajectory), this analysis quantifies the probability of reaching each of the trap-spaces identified in Figure~\ref{figure2}-A (i.e., the two stable states and the cyclic attractor), as shown in Figure~\ref{figure4}-B.
Under these conditions, about 50\% of the trajectories reach the cyclic attractor, versus about 45\% and 5\% for the two stable states. The convergence of these estimates as a function of the number of runs is displayed in Figure~\ref{figure4}-C. Interestingly, the first stable state (SS1, reached with a 45\% probability) corresponds to the \textit{bona fide} quiescent state, devoid of CycA activity in the absence of CycD input.

\begin{figure}[!t]
    \centering
    \includegraphics[width=0.9\linewidth]{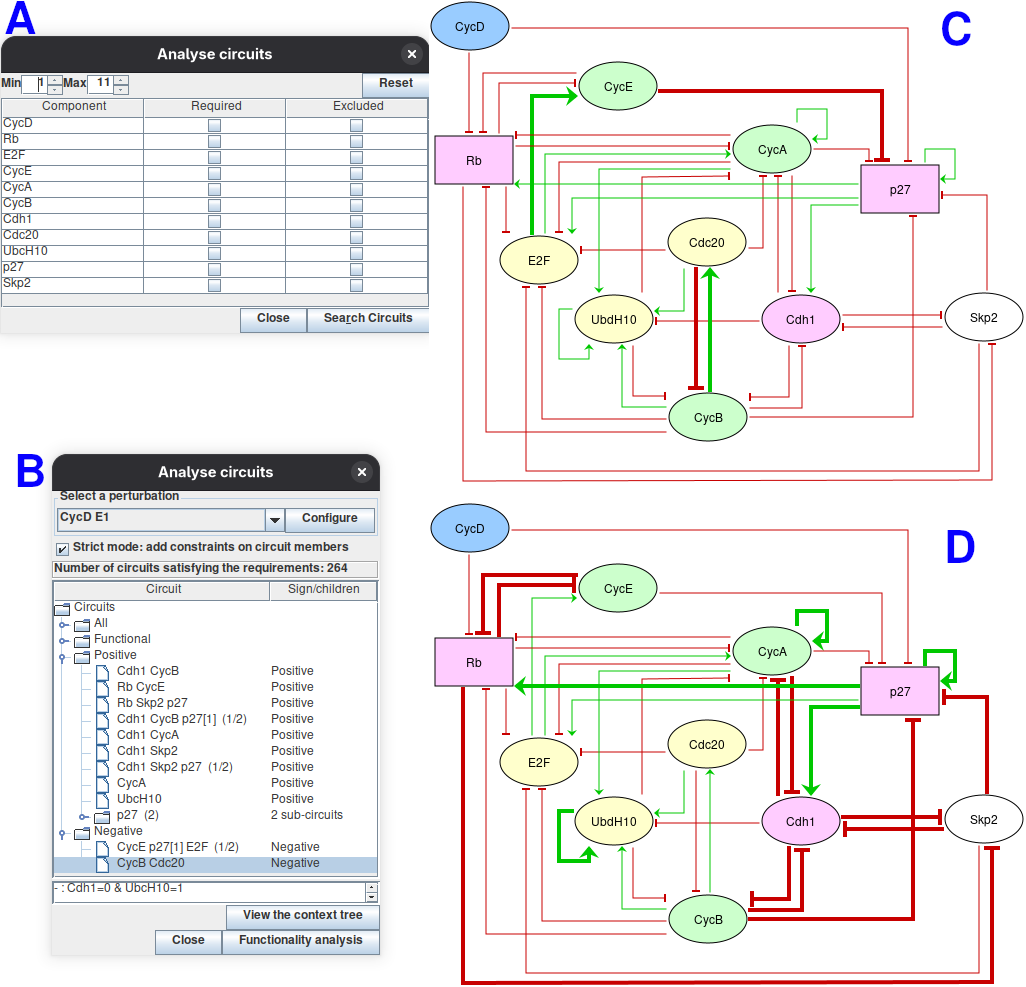}
    \caption{\textbf{Computation and visualisation of functional regulatory circuits.}
\textbf{A}: Interface enabling the identification of the regulatory circuits of a logical model, activated by selecting the \emph{Analyse circuits} functionality from the \emph{Tools} menu.
The user can restrict the size of the circuits to be returned (here considering circuits potentially involving from one component to the total number of components by default), as well as the inclusion or exclusion of specific components (not done for the model considered here, but very useful for large and complex models). 
\textbf{B}: Results obtained after pressing successively the \emph{Search Circuits} (panel A) and \emph{Functionality analysis} buttons, in the presence of the perturbation \emph{CycD E1} (i.e., for CycD ectopically activated). Circuits are listed according to their signs (positive or negative). 
GINsim identified two functional negative circuits, the first involving CycB and Cdc20, the second involving p27, E2F and CycE, plus ten functional positive circuits.
The negative circuit involving CycB and Cdc20 being selected, its context of functionality is displayed below (Cdh1 = 0 and UbcH10 = 1).
\textbf{C}: Regulatory graph with the two functional negative circuits emphasised with thicker arcs.
\textbf{D}: Regulatory graph with the ten functional positive circuits emphasised with thicker arcs.}
    \label{figure5}
\end{figure}

\paragraph{Identification of functional circuits}
One of the most remarkable functionalities of GINsim is the inclusion of an algorithm for the identification of the regulatory circuits present in a regulatory graph. Selecting \emph{Analyse circuits} in the \emph{Tools} menu activates the interface shown in Figure~\ref{figure5}-A.
Before running a search for circuits, the user can specify an interval of circuit lengths to consider or impose/exclude nodes, which is particularly useful when analysing very large and complex models.
Running the circuit search without any constraints for the mammalian cell cycle model leads to an interface listing the 264 circuits found in the regulatory graph, involving between one and ten nodes.
After selecting the \emph{CycD E1} perturbation, clicking on the \emph{Functionality analysis} button launches the interface shown in Figure~\ref{figure5}-B, listing two negative circuits and ten positive circuits, which are highlighted in Figure~\ref{figure5}-C and Figure~\ref{figure5}-D, respectively.

From this analysis, we can conclude that the functional negative circuits account for the observed oscillatory behaviour, while the positive functional circuits presumably drive irreversible decisions along the cell cycle, some of them corresponding to documented check-points.


\section{Conclusions}\label{sec:conclusions}
Since the launch of the first version of GINsim over 20 years ago, many tools have been developed to define Boolean models and analyse their dynamical behaviour, or to infer such models directly from gene expression data (see, e.g., \cite{Noel2025colomoto, Klemmer2026}, and references therein). In parallel, GINsim has been regularly maintained and extended, in particular to provide tools for the dynamical analysis of larger and more complex regulatory networks.
From our point of view, the current GINsim release combines the following assets, which to the best of our knowledge are not to be found in any other single tool:
\begin{itemize}
    \item A graphical interface enabling the delineation of regulatory graphs, the assignment of logical rules, and extensive documentation using free text and hyperlinks.
    \item The possibility to encode various model parametrisations in a single archive (.zginml file), combining information about single or multiple genetic perturbations, initial states, or patterns of interest. 
    \item Multiple modes of access, including a GUI, a command-line and scripting interface (e.g., Jython scripts, or Python within the CoLoMoTo notebook), and the sharing of the core Java libraries (bioLQM, MDDlib).
    \item The support of multilevel models and their automatic conversion into dynamically consistent Boolean models.
    \item A variety of functionalities to explore wild-type and mutant dynamical behaviours.
    \item A unique circuit functionality analysis tool.
\end{itemize}

We have illustrated the complementarity between selected tools and the resulting insights through their application to a moderate-size multilevel logical model for the core regulatory network controlling the progression of the mammalian cell cycle. Many more functionalities are described in GINsim online documentation. 

Finally, many (but not all) GINsim functionalities are available from the CoLoMoTo environment, developed by the Consortium for Logical Models and Tools (\url{https://colomoto.github.io/}) \cite{Naldi2018notebook,Noel2025colomoto}, which currently combines more than twenty modelling and analysis tools, made available through Conda packages and Docker containers. These tools can be called from Python Jupyter notebooks, thereby fostering sharing, reproducibility, and extension of sophisticated model analysis workflows.
In the future, we plan to extend the GINsim Python interface to provide access to more GINsim functionalities from the CoLoMoTo environment (e.g. regulatory circuit analysis, reachability analysis with Avatar).

\backmatter

\bmhead{Acknowledgements}
The authors acknowledge the key contributions of Aurélien Naldi to the development of previous releases of GINsim, as well as numerous constructive and insightful comments from GINsim users over the years, in particular, from Laurence Calzone, Élisabeth Remy and José Antonio Sánchez Villanueva.

\section*{Declarations}

\begin{itemize}
\item Funding:
This work was supported by national funds through FCT, under projects
UID/50021/2025 (DOI: 10.54499/UID/50021/2025),
FCT-Mobility investment 11/C06-i06/2024 (FCT/Mobility/1419033983/2024-25), by OSCARS project, funded by the European Union under the Horizon Europe grant No. 101129751, and by Excellence Laboratory Archimède (ANR-11-LABX-0033).
\item Code and Software availability:
GINsim (version 3.1) is open source under the GPL-3.0 license. Its code is available at \url{https://github.com/GINsim/GINsim}, and the GINsim website (\url{https://ginsim.github.io/install/}) provides the installation instructions.
\item Data availability:
The model used in this study is available in the GINsim model repository \url{https://ginsim.github.io/models/2016-mammal-cell-cycle/}.
\item Author contributions: PTM, DB, LT and OC were involved in the GINsim code. PTM and LT were involved in building the new GINsim website. DT and CC were involved in project supervision and securing funding. PTM, LT and DT drafted the manuscript. All authors read and approved the final manuscript.
\item Competing interests: The authors declare no competing interests.
\end{itemize}


\bibliography{sn-bibliography}

\clearpage

\begingroup 
\renewcommand{\thetable}{S\arabic{table}}\setcounter{table}{0}

\begin{table}[t]
\resizebox{\textwidth}{!}{
\begin{tabular}{l|p{9.5cm}|c}
\hline
\textbf{Nodes} & \textbf{Logical rules} & \textbf{Ref.} \\ \hline
CycD & Fixed input & \cite{Faure2006} \\ \hline
E2F & !Rb \& (!CycA $\mid$ (p27:1 \& !Skp2) $\mid$ p27:2) \& !(Cdc20 \& CycB) & \cite{Traynard2016} \\ \hline
CycE & E2F \& !Rb:2 & \cite{Faure2006} \\ \hline
CycA & ((E2F \& !Rb) $\mid$ CycA) \& (!UbcH10 $\mid$ (!Cdh1 \& !Cdc20)) & \cite{Traynard2016} \\ \hline
CycB & (!UbcH10 $\mid$ !Cdc20) \& !Cdh1 & \cite{Traynard2016} \\ \hline
Cdh1 & (!CycA $\mid$ (p27:1 \& !Skp2) $\mid$ p27:2) \& (!CycB $\mid$ p27) & \cite{Traynard2016} \\ \hline
Cdc20 & CycB & \cite{Faure2006} \\ \hline
UbcH10 & !Cdh1 $\mid$ (Cdh1 \& UbcH10 \& (CycA $\mid$ Cdc20 $\mid$ CycB)) & \cite{Faure2006} \\ \hline
Skp2 & !Cdh1 $\mid$ !Rb:2 & \cite{Traynard2016} \\ \hline
p27 $\Rightarrow$ 1 & CycD \& ((!CycB \& (!CycA $\mid$ p27) \& (!CycE $\mid$ (p27:2 \& !CycA))) $\mid$ !Skp2) & \cite{Traynard2016} \\ \hline
p27 $\Rightarrow$ 2 & !CycD \& ((!CycB \& (!CycA $\mid$ p27) \& (!CycE $\mid$ (p27:2 \& !CycA))) $\mid$ !Skp2) & \cite{Traynard2016} \\ \hline
Rb $\Rightarrow$ 1 & (CycD \& !CycE \& !CycA \& !CycB) $\mid$ (!CycD \& p27 \& CycE \& CycA \& CycB) $\mid$ (CycD \& (CycE $\mid$ CycA $\mid$ CycB) \& p27) & \cite{Traynard2016} \\ \hline
Rb $\Rightarrow$ 2 & (!CycD \& !CycE \& ((!CycA \& !CycB) $\mid$ p27)) $\mid$ (!CycD \& p27 \& CycE \& (!CycA $\mid$ !CycB)) & \cite{Traynard2016} \\ \hline
\end{tabular}}
    \caption{Logical rules for the multilevel logical model of the mammalian cell cycle, as defined by Traynard et al \cite{Traynard2016}. The rules combine the identifiers of the regulators with the operators NOT (!), AND (\&) and OR ($\mid$); in the case of a multi-level variable, X:n denotes that component X has a value of at least n. For the ternary nodes Rb and p27, two rules are given, enabling the value 1 ($\Rightarrow$ 1) and the value 2 ($\Rightarrow$ 2), respectively. The last column indicates the reference in which each rule was defined.}
    \label{tab:logical_rules}
\end{table}

\endgroup 

\end{document}